\documentclass[aps,prl,floatfix,twocolumn,showpacs,amsmath,amssymb,superscriptaddress]{revtex4}

\usepackage{bm,enumerate,amssymb}
\usepackage{comment,color}
\usepackage{graphicx}
\usepackage{subfigure}
\usepackage{hyperref}

\newcommand{\beq}{\begin{equation}}
\newcommand{\eeq}{\end{equation}}
\newcommand{\beqn}{\begin{eqnarray}}
\newcommand{\eeqn}{\end{eqnarray}}

\newcommand{\ua}{\uparrow}
\newcommand{\da}{\downarrow}
\newcommand{\ra}{\rightarrow}

\date{\today}

\begin{document}


\title{Interacting Topological Superconductors and
possible Origin of \\ $16n$ Chiral Fermions in the Standard Model}


\author{Yi-Zhuang You}

\author{Yoni BenTov}

\author{Cenke Xu}

\affiliation{Department of Physics, University of California,
Santa Barbara, California 93106, USA}

\begin{abstract}

Motivated by the observation that the Standard Model of particle
physics (plus a right-handed neutrino) has precisely 16 Weyl
fermions per generation, we search for $(3+1)$-dimensional chiral
fermionic theories and chiral gauge theories that can be
regularized on a 3 dimensional spatial lattice when and only when
the number of flavors is an integral multiple of 16. All these
results are based on the observation that local interactions
reduce the classification of certain $(4+1)$-dimensional
topological superconductors from $\mathbb{Z}$ to $\mathbb{Z}_{8}$,
which means that one of their $(3+1)$-dimensional boundaries can
be gapped out by interactions without breaking any symmetry when
and only when the number of boundary chiral fermions is an
integral multiple of $16$.

\end{abstract}


\maketitle

In the Standard Model (SM) of particle physics \footnote{We assume
the right-handed neutrino exists.}, when the energy scale is
higher than the vacuum expectation value (VEV) of the Higgs field,
$v = 246$ GeV, effectively there are in total 16 massless
left-handed chiral (Weyl) fermions in each generation:
\begin{equation}
H = \int d^3x \sum_{n\,=\,1}^3\sum_{a\,=\,1}^{16} \psi_{a,n}^\dagger \,i\vec\sigma \cdot \vec\partial\, \psi_{a,n}\;.
\end{equation}
For example, in the first generation ($n = 1$) there are left
chiral fermions $(u_\alpha, d_\alpha)_L$, $(u^\dagger_\alpha,
d^\dagger_\alpha)_R$, $(e, \nu_e)_L$, $(e^\dagger,
\nu^\dagger_e)_R$~\footnote{Under complex conjugation
(particle-hole transformation) the right chiral fermions become
left fermions.}, where $\alpha = 1,2,3$ is the color index. In the
SO(10) Grand Unified Theory (GUT)~\cite{SO10A,SO10B,SO10C}, these
fermions couple to the SO(10) gauge field as a 16 dimensional
irreducible spinor representation of the SO(10) gauge group, which
we will denote by $\psi \sim 16_+$. Since both the SM and the
SO(10) GUT have no known gauge anomaly, it is expected that they
can be regularized as a full quantum system on a three dimensional
spatial lattice. However, it is well-known that if we want to
write down a lattice model for the 16 left fermions in the SM with
chiral gauge couplings, we will inevitably also obtain ``mirror"
fermions, $\psi' \sim 16_-$ (16 right fermions) in the low-energy
effective field theory that arises from the same lattice
model~\cite{doublingA,doublingB}.

In order to get around the fermi doubling theorem, one method is
to realize the SM on the $3d$ boundary of a $4d$ topological
insulator (TI) or topological superconductor
(TSC)~\cite{domainwall,kaplan1992,kaplan2012}. Then the 16 mirror
right Weyl fermions will be localized on the other opposite
boundary, which is spatially separated from the SM. Fermions at
each boundary can naturally have a chiral coupling to the bulk
gauge fields. However, this method requires subtle adjustment of
the scale of the fourth dimension: if the fourth dimension is too
large, the gauge boson in the bulk will be gapless and interfere
with the low energy physics of the boundary; on the other hand if
the fourth dimension is too small, then the SM suffers from
interference with its mirror sector on the other
boundary~\cite{latticefermions}.

It would be ideal if we can gap out the mirror sector without
affecting the left fermions in the SM. Then we can regularize the
SM on the $3d$ boundary of a $4d$ TI or TSC with a very thin
fourth dimension (which makes the bulk generically a $3d$ system).
However, if the mirror sector is gapped out in the standard way,
namely they are gapped out by condensing a boson field that
couples to the Majorana mass (Cooper pair mass) of the right
fermions, $\psi'^t_a i\sigma^y \psi'_b$, then the same boson field
would couple to the left fermions and gap them out as well. Thus
we seek a new mechanism to gap out the 16 right fermions with
interactions, such that the mirror fermions are gapped while
having zero bilinear expectation value, $\langle \psi_a'^t
i\sigma^y \psi'_b\rangle = 0$, for arbitrary flavor indices $a,b$
$= 1,...,16$. We will now drop the primes for notational
convenience, and it should be understood that when we attempt to
gap out the fermions $\psi_a$ we intend to gap out the mirror
sector while leaving the ordinary fermions gapless.

\begin{figure}
\begin{center}
\includegraphics[width=230pt]{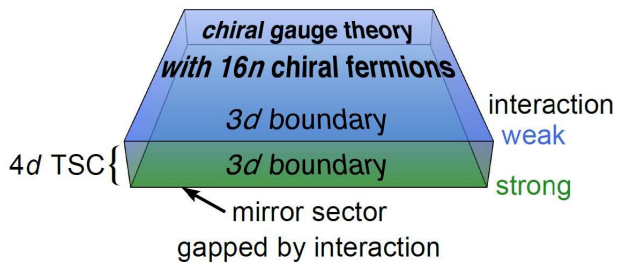}
\caption{ Our proposal of regularizing SM (or other anomaly free
chiral gauge theories with $16n$ chiral fermions) on the lattice.
The chiral gauge theory will be realized on the $3d$ boundary of a
$4d$ topological superconductor with a thin fourth dimension
(which makes the bulk a $3d$ system), and the mirror sector is
realized on the opposite boundary. Without interaction, the
symmetry forbids any fermion bilinear mass at the boundary; but
interaction can gap out the mirror sector without generating a
fermion bilinear mass term. Thus the only low energy degree of
freedom is the chiral gauge theory on the top boundary. However,
{this is only possible with $16n$ chiral fermions on each
boundary.}} \label{4dbulk}
\end{center}
\end{figure}

At this point we would like to make clear that we do not yet have
a lattice regularization for the full SM. Nevertheless, we do
provide explicit examples of non-abelian chiral gauge theories
where the above condition is satisfied if and only if they have
$16n$ Weyl fermions, and we hope that this novel result strongly
re-invigorates the search for a lattice regularization of the SM.

The new mechanism described in the previous paragraph becomes
possible if the bulk $4d$ state satisfies the following criteria:

{\it 1.} The interacting bulk system should {\it not} have the
global charge U(1) symmetry $\psi_a \rightarrow e^{i\theta}
\psi_a$, since this symmetry is anomalous once the boundary chiral
fermions are coupled to the gauge fields in the SM. Thus according
to the anomaly matching
condition~\cite{anomalymatching,anomalymatching2}, if the boundary
chiral fermions can be gapped by interaction, this system
necessarily breaks this U(1) symmetry. This implies that the bulk
system is a topological superconductor instead of topological
insulator.

{\it 2.} Each $3d$ boundary of the $4d$ TSC has $k$ chiral Weyl
fermions ($k$ is a divisor of 16), and the $4d$ TSC has a
$\mathbb{Z}$ classification without interaction, namely for
arbitrary flavor numbers, the symmetries that define the bulk TSC
forbid any fermion bilinear mass $m_{ab}\, \psi^t_{a} i\sigma^y
\psi_{b} + H.c.$ at the $3d$ boundary (here $m_{ab}$ is a
symmetric matrix). In other words at the {\it noninteracting}
level the $3d$ boundary cannot be gapped without breaking
symmetry, for arbitrary flavor number.

{\it 3.} Under interaction, the classification of the $4d$ TSC is
reduced to $\mathbb{Z}_{16/k}$, namely when the boundary has 16
flavors of $(3+1)d$ chiral fermions, local interactions can gap
out the boundary without spontaneously breaking any symmetry. In
other words, the boundary chiral fermions can be gapped by
interaction without spontaneously generating a fermion bilinear
mass term, $i.e.$ $\langle \psi^t_{a} i\sigma^y \psi_{b} \rangle =
0$.

Notice that there must be a minimum nonzero critical interaction
strength for this gapping mechanism to work, because a weak
short-range four fermion interaction is irrelevant for $(3+1)d$
Dirac or chiral fermions. Thus we assume that the interaction on
one boundary is stronger than the other, thus the interaction only
gaps out one of the boundaries.

Such interaction reduced classification of TSC has been studied in
theoretical condensed matter physics in recent years. The
pioneering work was done by Fidkowski and Kitaev. They
demonstrated explicitly that a $1d$ TSC with time-reversal
symmetry, which in the noninteracting limit has a $\mathbb{Z}$
classification~\cite{ludwigclass1,ludwigclass2,kitaevclass}, has
only a $\mathbb Z_8$ classification in the presence of local
interactions~\cite{fidkowski1,fidkowski2}. Each copy of this $1d$
TSC will lead to one Majorana zero mode $\gamma_a$ at its $0d$
boundary. These Majorana zero modes transform trivially under
time-reversal ($Z_2^T$): \beqn Z_2^T: \gamma_a \rightarrow
\gamma_a. \eeqn Since time-reversal is antiunitary, this symmetry
guarantees that no fermion bilinear term is allowed at the
boundary, $i.e.$ all terms of the form $i\gamma_a\gamma_b$ are odd
under $Z_2^T$. Thus without interaction, the boundary of this $1d$
TSC is always degenerate, although the bulk is gapped and
nondegenerate. However, Fidkowski and Kitaev demonstrated that a
time-reversal invariant four-fermion interaction can gap out the
$0d$ boundary without spontaneously breaking the $Z_2^T$ (the $0d$
boundary is gapped with $\langle i\gamma_a \gamma_b \rangle = 0$),
when and only when the system has $8n$ copies of such a $1d$ TSC.
This implies that the $1d$ time-reversal symmetry protected TSC
has only $\mathbb{Z}_8$ classification under interaction.


The work in Ref.~\cite{fidkowski1,fidkowski2} was soon generalized
to $2d$ TSC with a $1d$
boundary~\cite{qiz8,yaoz8,zhangz8,levinguz8}. For example,
Ref.~\cite{qiz8} studied a $2d$ TSC with both $Z_2$ and
time-reversal symmetry, whose gapped bulk is simply a $p \pm ip$
TSC with $p_x + i p_y$ pairing for spin-up fermion $c_\uparrow$,
and $p_x - ip_y$ pairing for spin-down fermion $c_\da$. The $1d$
boundary of this TSC has a $1d$ nonchiral Majorana fermion with
Hamiltonian: \beqn H = \int dx \ ( \chi_L i
\partial_x \chi_L - \chi_R i \partial_x \chi_R ). \label{1db} \eeqn
$\chi_L$ and $\chi_R$ are Bogoliubov quasiparticles of $c_{\ua}$
and $c_\da$ respectively. On the $1d$ boundary the $Z_2$ and
$Z_2^T$ transformations act as the following: \beqn Z_2 : \chi_L
\rightarrow \chi_L, \ \ \chi_R \rightarrow - \chi_R, \cr \cr Z_2^T
: \chi_L \ra \chi_R, \ \ \chi_R \ra \chi_L. \label{1dsym} \eeqn
With these symmetries, it is straightforward to verify that for
arbitrary numbers of the boundary Eq.~\ref{1db}, any fermion
bilinear mass term is forbidden. For example, $\bar{\chi} \chi = 2
i \chi_L \chi_R $ is forbidden by the $Z_2$ symmetry.
Ref.~\cite{qiz8} showed that although all the fermion bilinear
mass terms are forbidden by symmetry at the boundary, when there
are $8n$ copies of this $p \pm ip$ TSC, a particular four fermion
interaction term which preserves both $Z_2$ and $Z_2^T$ will still
gap out the boundary without degeneracy, namely the $1d$ boundary
is gapped with $\langle \bar{\chi}_a \chi_b \rangle = 0$ for
arbitrary flavor index $a,b$.

Ref.~\cite{chenhe3B,senthilhe3} studied the classification of $3d$
TI/TSC under interaction. For example, Ref.~\cite{senthilhe3}
studied a $3d$ TI with U(1) and time-reversal symmetry whose $2d$
boundary is described by the Hamiltonian \beqn H = \int d^2x \
\psi^\dagger (i \sigma^x
\partial_x + i \sigma^z \partial_y) \psi. \label{2db} \eeqn The U(1)
and time-reversal symmetry act as: \beqn U(1) : \psi \rightarrow
e^{i\theta} \psi, \ \ \  Z_2^T : \psi \rightarrow \sigma^y
\psi^\dagger, \label{2dsym}\eeqn which forbid all the fermion
bilinear mass terms at the $2d$ boundary for arbitrary copies of
the system, namely the classification at the noninteracting level
is $\mathbb{Z}$. Nevertheless, Ref.~\cite{senthilhe3} argued that
a U(1) and $Z_2^T$ invariant short range interaction reduces the
classification of this $3d$ TI to $\mathbb{Z}_8$, namely when
there are $8n$ copies of Eq.~\ref{2db} at the $2d$ boundary,
interaction can gap out the boundary without spontaneously
generating any fermion bilinear mass term. Ref.~\cite{senthilhe3}
argued that many $3d$ TIs and TSCs with different symmetries have
similar interaction-reduced classifications, and it is almost
universally true that when the $2d$ boundary has 16 $2d$ Majorana
fermions, the boundary can be trivially gapped out by interaction.

In this work we will generalize the works summarized above to four
spatial dimensions. Before making connection to the SM, we will
first study a simple example of $4d$ TSC whose classification is
reduced by interaction. In particular we will study a $4d$ TSC
whose boundary contains two flavors of $(3+1)d$ chiral fermions:
\beqn H = \int d^3x \ \sum_{a = 1}^2 \psi^\dagger_a (i
\vec{\sigma} \cdot \vec{\partial}) \psi_a. \label{3db} \eeqn We
define the following U(1), $Z_2$ and time-reversal symmetry on
$\psi_a$: \beqn U(1) &:& \psi_a \rightarrow [e^{i \tau^y
\theta}]_{ab} \psi_b, \cr\cr Z_2 &:& \psi_a \rightarrow
(\tau^y)_{ab} \psi_b, \cr\cr Z_2^T &:& \psi_{a} \rightarrow K
\sigma^y (\tau^y)_{ab} \psi_b, \label{sym} \eeqn where $K$ is a
complex conjugation. For the bulk state, we can use the same bulk
band structure introduced for the $4d$ quantum Hall state in
Ref.~\cite{qi2008,wen2013}: \beqn H &=& \sum_{a = 1}^2 \sum_k
\psi^\dagger_{k,a} \left( \sum_{i = 1}^4 \Gamma_i \sin(k_i)
\right) \psi_{k,a} \cr\cr &+& m \ \psi^\dagger_{k,a} \Gamma_5
\left( \sum_{i = 1}^4\cos(k_i) - 3\right) \psi_{k,a} , \eeqn Where
$\Gamma_{1,2,3} = \sigma^{1,2,3} \otimes \rho^3 $, $\Gamma_{4} =
\mathbf{1}_{2\times 2} \otimes \rho^2$, $\Gamma_{5} =
\mathbf{1}_{2\times 2} \otimes \rho^1$, where $\rho^a$ are another
set of Pauli matrices. The $3d$ boundary of this theory has
precisely two flavors of chiral fermions Eq.~\ref{3db}.

As long as we preserve the $U(1) \times Z_2 \times Z_2^T$
symmetry, the $3d$ boundary can never be gapped without
interaction for arbitrary copies of this system, because the only
fermion bilinear mass terms that can gap out the boundary are the
Cooper pair operators: $\psi^t_a i \sigma^y \psi_b + H.c.$ which
inevitably break at least one of the symmetries. Thus this $4d$
TSC has a $\mathbb{Z}$ classification with the $U(1) \times Z_2
\times Z_2^T$ symmetry, at the noninteracting level.

In the following we will argue that short range interactions can
reduce the classification of this $4d$ TSC to $\mathbb{Z}_{8}$:
local four-fermion interactions can gap out 8 copies of
Eq.~\ref{3db} (i.e. 16 chiral fermions at the $3d$ boundary)
without generating a nonzero expectation value for any fermion
bilinear mass operator. Since a weak short range interaction is
irrelevant, the interaction has to have a magnitude larger than
some nonzero critical value in order to successfully gap out the
fermions.

Directly studying strong four-fermion interactions is difficult,
so we will follow the same logic as in Ref.~\cite{senthilhe3}: we
will first manually break a subgroup of the $U(1) \times Z_2
\times Z_2^T$ symmetry by condensing an order parameter that
transforms nontrivially under these symmetries. Then we will
proliferate the defects of the condensate to restore the broken
symmetry. After proliferating the defects, the order parameter
becomes disordered and can be safely integrated out. This
generates an effective four-fermion interaction at low energy.

The nature of the phase after proliferating the defects depends on
the quantum numbers and spectrum of the defects. \textit{The
desired fully symmetric, gapped and nondegenerate state is only
possible when the defects in the condensate have a trivial
spectrum.} We will analyze three different types of order
parameters and defects, and all these defects suggest that a
symmetric, fully gapped, and nondegenerate boundary state is only
possible when there are 16 chiral fermions at the $3d$ boundary.

\begin{figure}
\begin{center}
\includegraphics[width=245pt]{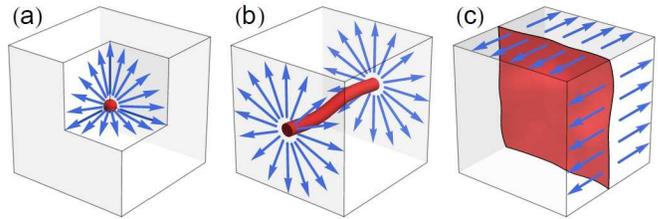}
\caption{Illustration of topological defects: (a) monopole, (b)
vortex line, (c) domain wall. Without interaction, all these
defects are nontrivial, i.e. they have degenerate/gapless spectra.
However, interactions make all these defects gapped and
nondegerate, thus after proliferating these defects, the $3d$
boundary enters a symmetric, fully gapped and nondegenerate phase.
} \label{fig: topological defects}
\end{center}
\end{figure}

Let us first spontaneously break the U(1) symmetry by condensing
an O(2) ``superfluid" order parameter at the $3d$ boundary: \beqn
\vec{\phi} = (\mathrm{Re}[\psi^t \sigma^y \tau^x \psi], \ \
\mathrm{Re}[\psi^t \sigma^y \tau^z \psi]). \label{o2} \eeqn This
superfluid order parameter gaps out the chiral fermions and breaks
the U(1) symmetry, but preserves the $Z_2$ and $Z_2^T$ symmetries
in Eq.~\ref{sym}. The broken U(1) symmetry can be restored by
proliferating the vortex lines of the O(2) order parameter in
Eq.~\ref{o2} (Fig.~\ref{fig: topological defects}$b$).
Proliferation of the vortex line can be systematically described
in the dual formalism. However, we have to be careful with the
core of the vortex line, since it is the singularity of the O(2)
order parameter, and the fermions may become gapless along the
vortex line. In our current case, the vortex line of this O(2)
order parameter traps $1d$ nonchiral Majorana fermion modes that
are localized at the vortex line. Upon solving the Dirac equation
in the vortex background, we find that these modes are described
by the Hamiltonian of Eq.~\ref{1db}, and their transformation
properties under the residual $Z_2$ and $Z_2^T$ symmetries are
precisely those given in Eq.~\ref{1dsym}. Thus as we already
argued, without interaction, this $1d$ system cannot be gapped
without degeneracy, for arbitrary copies of this system.

However, based on the results in
Ref.~\cite{qiz8,yaoz8,zhangz8,levinguz8}, we know that for $8n$
copies of this system, a $Z_2$ and $Z_2^T$ invariant short range
interaction at the vortex line can gap out these $1d$ Majorana
modes without degeneracy, $i.e.$ without spontaneously breaking
the $Z_2$ and $Z_2^T$ symmetry. \textit{Thus when and only when
the $3d$ boundary has 16 chiral fermions can we gap out the
boundary without generating a fermion bilinear mass term.}

We can also analyze other different types of defects of the
system. For example, we can temporarily break the $Z_2$ symmetry
in Eq.~\ref{sym} by condensing the Ising order parameter \beqn
\phi = \mathrm{Im}[\psi^t \sigma^y \psi], \label{z2} \eeqn which
only breaks the $Z_2$ symmetry but preserves the U(1) and $Z_2^T$.
The domain wall of the $Z_2$ order is a $2d$ manifold
(Fig.~\ref{fig: topological defects}$c$), and presumably
proliferating the domain wall can restore the $Z_2$ symmetry.
However, just like the previous paragraph, the $Z_2$ domain wall
may carry gapless fermion modes. By solving the Dirac equation at
the domain wall directly, we can see that there are two flavors of
gapless $2d$ Majorana fermions described by Eq.~\ref{2db} that are
localized at the domain wall. These $2d$ Majorana fermions have
the residual $U(1) \times Z_2^T$ symmetry, which act on the domain
wall states precisely in the same way as Eq.~\ref{2dsym}. And as
long as the $U(1) \times Z_2^T$ symmetry is preserved, all the
fermion bilinear mass terms at the domain wall are forbidden, for
arbitrary copies of the systems. Thus without interaction at the
domain wall, proliferation of the $2d$ domain wall will not lead
to a trivially gapped phase. However, it was demonstrated in
Ref.~\cite{senthilhe3} that 8 copies of Eq.~\ref{2db} (i.e. 8
copies of our current system) with $U(1) \times Z_2^T$ symmetry
can be gapped out without degeneracy by local interactions. Thus
once again a $3d$ system with 16 chiral fermions becomes special:
\textit{when there are 16 chiral fermions at the $3d$ boundary, we
can obtain a fully gapped nondegenerate state by proliferating the
$Z_2$ domain walls.}

The third type of scenario we consider is an ordered phase at the
$3d$ boundary that breaks both U(1) and $Z_2$ symmetry, while
keeping the $Z_2^T$ symmetry unbroken. In this phase both the O(2)
vector and $Z_2$ order parameters in the previous two scenarios
condense. The defect that can restore all the symmetries after its
proliferation is a ``hedgehog" like monopole (Fig.~\ref{fig:
topological defects}$a$), which is a vortex line penetrating the
$Z_2$ domain wall. Based on the results in Ref.~\cite{jackiw1976},
there is a Majorana fermion zero mode at the core of the hedgehog
monopole, which transforms trivially under time-reversal symmetry,
$Z_2^T: \gamma \rightarrow \gamma$. Based on the results in
Ref.~\cite{fidkowski1,fidkowski2}, we know that with 8 copies of
the system Eq.~\ref{3db}, a local interaction that preserves the
$Z_2^T$ interaction can gap out all the 8 Majorana zero modes at
the monopole without degeneracy. Then condensing the monopoles can
restore all the symmetries and lead to a fully gapped and
nondegenerate state at the $3d$ boundary.

In the analysis above we argued that 8 flavors of the $4d$ TSC
with 16 chiral fermions at its $3d$ boundary become trivial under
interaction. In fact this interaction can have a much larger
symmetry than the assumed $U(1) \times Z_2 \times Z_2^T$. For
example, the U(1) and $Z_2$ order parameters considered previously
(Eq.~\ref{o2}, \ref{z2}) can be combined together to form an SU(2)
vector, \beqn \vec\phi = \text{Re}\!\left[\psi^t (\sigma^y\otimes
i\tau^y\vec\tau\,)\psi\right]\;.\eeqn The four-fermion interaction
that guarantees the triviality of the hedgehog monopole of this
$SU(2)$ vector can have a global symmetry at least as large as
$SO(7)$, under which the 8 flavors of $4d$ TSC transform as an
8-dimensional real spinor representation. This $SO(7)$ symmetry
contains the obvious subgroup $SO(6)$, which is locally isomorphic
to $SU(4)$, and this in turn contains the obvious subgroup
$SU(3)$. Under a hypothetical breaking of $SO(7) \to SU(4)$, the
spinor representation would decompose as $8 \to 4\oplus\bar 4$.

In the analysis above we have ignored the gauge fields. We assume
that the four fermion interactions play the main role in gapping
out the fermions. We can ``gauge" the global symmetries of the
$4d$ TSC by coupling the fermions to the gauge fields. For
example, we can couple 8 flavors of the $4d$ TSC discussed in this
paper to $SO(7) \times SU(2)$ gauge fields, and one of the $3d$
boundaries is gapped out by short range interactions. Since the
boundary fermions are gapped out without generating any fermion
bilinear mass or breaking any symmetry, the gauge fields will {\it
not} be massive due to Higgs mechanism.
As long as the scale of the fourth dimension is small enough, the
only low energy degrees of freedom left are 16 chiral fermions on
the other boundary coupled to $SO(7) \times SU(2)$ gauge fields.
This theory has no gauge anomaly, but as far as we know, there is
no other simple way of realizing this chiral gauge theory on a
$3d$ lattice. For example, in a system called ``Weyl
semimetal"~\cite{weylsemimetal}, we can realize 8 left and 8 right
chiral fermions at different momenta in the $3d$ Brillouin zone.
After a particle-hole transformation, right Weyl fermions will
become left fermions, but there would be no $SO(7) \times SU(2)$
continuous symmetry that mixes the chiral fermions at different
momenta. Thus the 16 chiral fermions realized in this conventional
way cannot be coupled to $SO(7) \times SU(2)$ gauge fields on the
lattice. The new mechanism studied in this paper provides a
regularization of this chiral gauge theory.

Our results suggest a possible direction of realizing the SM on a
$3d$ lattice. If we want to realize the SM in the same way, $i.e.$
realize the SM on one boundary of a $4d$ TSC and gap out its
mirror sector by local interactions, then we need to argue that
the classification of the $4d$ TSC with $SU(3)\times SU(2) \times
U(1)$ symmetry is reduced by short range interactions. We do not
have a full argument for this conclusion yet. However, this
conclusion is consistent with the fact that the SM with a
right-handed neutrino has 16 chiral fermions in each generation.
Recently, Ref.~\cite{wen2013} has also proposed that the mirror
sector of the SM can be gapped out by interactions, but the
importance of the flavor number 16 was not pointed out. Based on
the observation in this paper, we conclude that the mirror sector
can be gapped out by interaction without interfering with the SM
only when there are $16$ chiral fermions per generation.

The authors are supported by the the David and Lucile Packard
Foundation, Sloan Foundation, and NSF Grant No. DMR-1151208. The
authors thank A. Zee for very helpful discussions.


\bibliography{Z16}

\end{document}